\documentclass[conference]{IEEEtran}
\usepackage{amsfonts}
\IEEEoverridecommandlockouts

\ifCLASSINFOpdf
\else
\fi
\usepackage{epsfig}
\usepackage{graphicx}
\usepackage{subfigure}
\usepackage{psfig}
\usepackage{booktabs}
\usepackage{epsf}
\usepackage{bm} 
\usepackage{slashbox}
\usepackage[cmex10]{amsmath}
\usepackage{amsthm}
\usepackage{booktabs}
\usepackage{color}

\begin{document}
\title{An Improved and More Accurate Expression for a PDF Related to Eigenvalue-Based Spectrum Sensing}
\author{\IEEEauthorblockN{Fuhui Zhou, \emph{Member, IEEE}, and Norman C. Beaulieu, \emph{Fellow, IEEE}}
\thanks{Fuhui Zhou is with the School of Information Engineering, Nanchang University, P. R. China, 330031 (e-mail: zhoufuhui@ncu.edu.cn).

Norman C. Beaulieu is with Beijing University of of Posts and Telecommunications and the Beijing Key Laboratory of Network System Architecture and Convergence, Beijing, China 100876 (e-mail: nborm@bupt.edu.cn).
}
\thanks{The research was supported by the National Natural Science Foundation of China (61701214), the Young Natural Science Foundation of Jiangxi Province (20171BAB212002), and the China Postdoctoral Science Foundation (2017M610400).}}
\maketitle
\begin{abstract}
Cooperative spectrum sensing based on the limiting eigenvalue ratio of the covariance matrix offers superior detection
performance and overcomes the noise uncertainty problem. While an exact expression exists, it is complex and multiple useful approximate expressions have been published in the literature. An improved, more accurate, integral solution for the probability density function of the ratio is derived using order statistical analysis to remove the simplifying, but incorrect, independence assumption. Thereby, the letter makes an advance in the rigorous theory of eigenvalue-based spectrum sensing.
\end{abstract}
\IEEEpeerreviewmaketitle
\section{Introduction}
\IEEEPARstart{T}{HE} Eigenvalue-based detection schemes, using the eigenvalues of the covariance matrix to construct a test statistic, are considered to be one of the most effective methods to test for the presence of a primary user (PU) signal in cognitive radio systems \cite{Y. Zeng}. The maximum-to-minimum eigenvalue (MME) detector is based on the ratio of the largest eigenvalue to the smallest eigenvalue of the covariance matrix. However, theoretical results for eigenvalue ratio schemes usually depend on asymptotic assumptions, since the distribution of the ratio of two extreme eigenvalues is difficult to compute \cite{Y. Zeng}-\cite{Shakir}.

The probability of false alarm (PFA) is the probability that the PU is absent but is detected to be present. PFA is one of the most important performance metrics in spectrum sensing for cognitive radio. Accurate determination of the PFA improves the accuracy of the decision threshold of a detector, and the efficiency of spectrum utilization. Note that the derivation of the PFA is dependent on the probability density function (PDF) of the test statistic of the detector under the hypothesis that the PU is absent. Meanwhile, this PDF is not known exactly in a tractable form suitable for further theoretical analysis, and the derivation of a popular approximation necessarily employs assumptions that are not rigorously correct mathematically. This letter makes a contribution to the theory of eigenvalue-based spectrum sensing in two ways, by removing invalid assumption, and by deriving a more accurate mathematically tractable approximation to this important PDF after removing the invalid assumption.

The distribution of the ratio of the two extreme eigenvalues for the MME scheme is commonly approximated by the Tracy-Widom distribution based on the Tracy-Widom law \cite{Y. Zeng}, \cite{F. Penna}. However, this approximation is based on the unrealistic assumption that the number of the received signal samples as well as the number of the cooperating secondary users are infinite. It has been shown that this approximation is poor when the number of the signal samples is small \cite{Shakir}, \cite{Shakir2}. An improved approximation to the PDF of the ratio of the two extreme eigenvalues is derived in \cite{Shakir} and \cite{Shakir2} with the assumption that the two extreme eigenvalues are independent and Gaussian distributed. In \cite{Matthaiou} and \cite{F. Penna2}, an exact PDF for the ratio of the two extreme eigenvalues has been derived. The derived exact expression for the PDF is quite complex, and other authors have chosen to use the approximation over the exact solution in \cite{Shakir}, \cite{Shakir2}, \cite{F. H. Zhou1}-\cite{Sharma}.

In this paper, an improved PDF approximation of the ratio of the two extreme eigenvalues is derived by using the order statistic theory and the Gaussian distribution assumption. It is shown that the derived PDF is more accurate than the commonly employed PDFs given in \cite{F. Penna} and \cite{Shakir}, and is simpler than the exact PDF given in \cite{Matthaiou}.

The rest of this paper is organized as follows. Section II presents the system model and MME spectrum sensing. An improved solution for the PDF of the ratio of two extreme eigenvalues is presented in Section III. Section IV presents simulation results. The paper concludes with Section V.

\section{System Model and MME Spectrum Sensing}
\begin{figure}[!t]
\centering
\includegraphics[width=3.5 in]{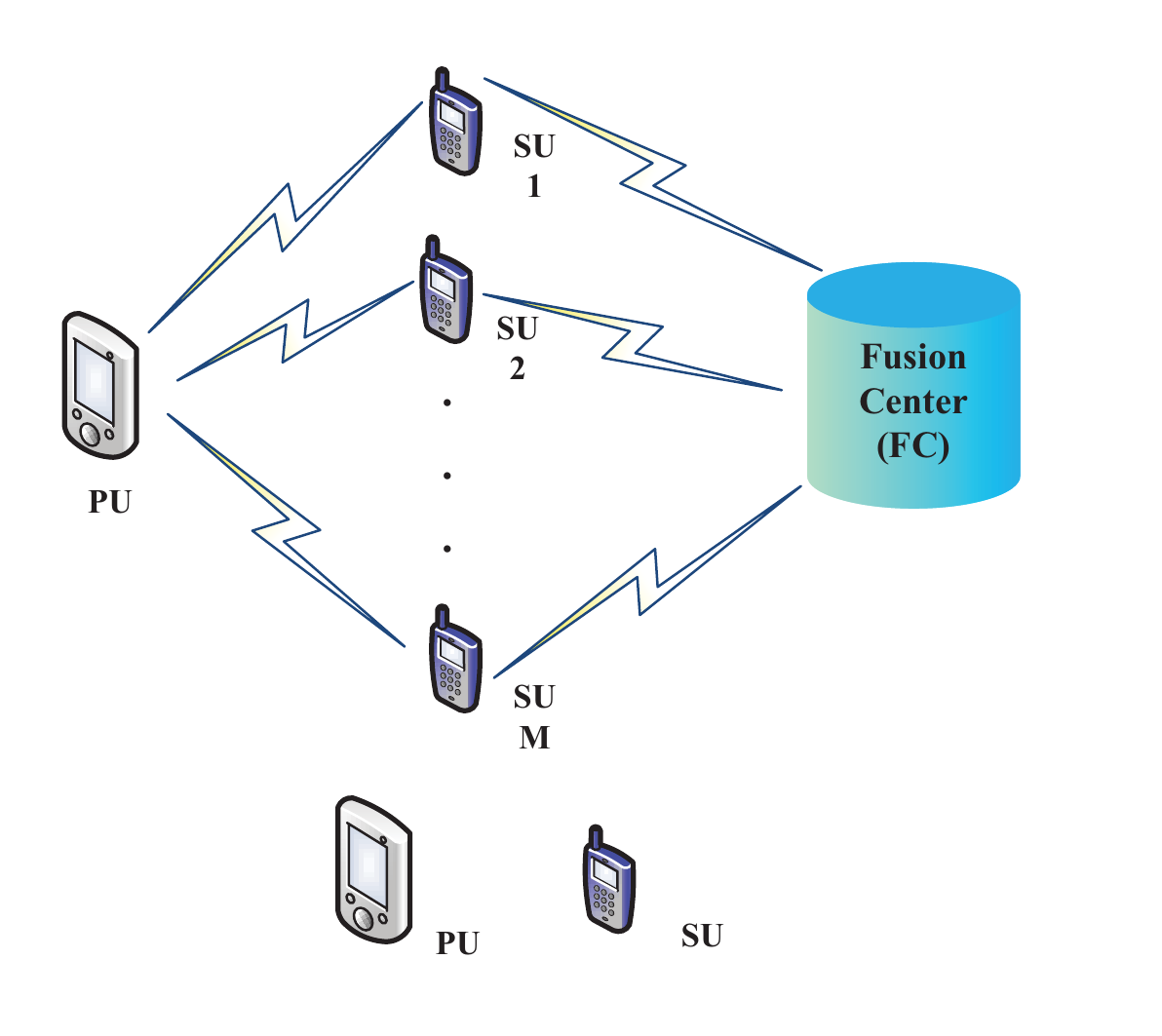}
\caption{The system model.} \label{fig.1}
\end{figure}

As shown in Fig. 1, a cognitive radio network with $M$ SUs that cooperatively detect one PU is considered. During the sensing time, each SU collects $N$ samples of received signal, denoted by ${{x_i}\left( n \right)}$, where $n = 1 , 2, \cdots, N$ and $i = 1, 2 , \cdots, M$.  Note that all the SUs receive the signal at the same time.  To
achieve synchronous sampling, each SU has the center frequency derived from the local oscillator and the same digital clock \cite{Y. Zeng}, \cite{F. Penna}, \cite{Shakir}, \cite{Shakir2}. This system model for cooperative spectrum sensing has been widely applied in these works. Then, the samples are transmitted to the fusion center. The aim of cooperative spectrum sensing is to construct a test statistic and make a decision between the two hypotheses $\left( {H_0} \ \textrm{and}  \ {H_1} \right)$ based on those collected samples, where $H_0$ denotes the absence of the PU, and $H_1$ represents the presence of the PU. Thus, the samples from each SU under the two hypotheses are given as
\begin{subequations}
\begin{align}
&{{H_0}:}\ {{x_i}\left( n \right) = {w_i}\left( n \right)} \\
&{{H_1}:}\ {{x_i}\left( n \right) = {h_i}\left( n \right)\sqrt P_s {s}\left( n \right) + {w_i}\left( n \right)}
\end{align}
\end{subequations}
where ${w_i}\left( n \right)\sim {\mathcal{CN}}\left( {0,\sigma _w^2 }\right)$ is complex Gaussian noise and $\mathcal{ CN}\left( {0,\sigma _w^2 } \right)$ denotes the complex Gaussian normal distribution with mean zero and variance $\sigma _w^2$. ${s}\left( n \right)$ is the primary user signal and ${h_i}\left( n \right)$ are the channel coefficients. $P_s$ is the transmitted power of the primary user. The distribution of the PU signal is unknown and independent of the noise. Based on the collected samples from $M$ SUs, a data matrix is defined as $\mathbf{X} = \left[ {X_1^T, X_2^T, \cdots , X_M^T} \right]$, where ${X_m} = \left[{{x_m}\left( 1 \right)}, \ {{x_m}\left( 2 \right)},\  \cdots, \ {{x_m}\left( N \right)}\right]$ with $m=1, 2, \cdots, M$. The sample covariance matrix is defined as ${\mathbf{R}_x} = \left( {1/N}\right) \mathbf{X}{\mathbf{X}^H}$, where ${\left( \cdot \right)^H}$ represents the Hermitian transpose operator. Let ${\lambda _{1}} \ge {\lambda _{2}} \ge  \cdots  \ge {\lambda _{M}}$ denote the ordered eigenvalues of the matrix ${\mathbf{R}_x}$. The test statistic for the MME spectrum sensing scheme was formulated in \cite{Shakir}. It is denoted by $T_\xi$, given as
\begin{align}\label{27}\
\begin{split}
{T_\xi }  = \frac{{\lambda _1 }}{{\lambda _M }}\mathop  {\mathbin{\lower.4ex\hbox{$\buildrel{\mathbin{\buildrel\scriptstyle<\over
{\smash{\scriptstyle\relbar}\vphantom{_{\scriptstyle x}}}}}\over
{\smash{\scriptstyle>}\vphantom{_x}}$}}} \limits_{H_1 }^{H_0 } {\gamma _\xi }
\end{split}
\end{align}
where ${\gamma _\xi }$ is the decision threshold of the MME spectrum sensing scheme.
\section{Improved Solution for The PDF of The Ratio of Two Extreme Eigenvalues Under Hypothesis $H_0$}
In \cite{Y. Zeng}, \cite{F. Penna}, the PDF of the ratio of two limiting eigenvalues under the hypothesis $H_0$ is approximated by the PDF of the Tracy-Widom distribution. This approximation is poor when the number of samples is small or moderate. The PDF of the ratio is derived based on the assumption that the two limiting eigenvalues are independent normal random variables \cite{Shakir}. The assumption that the two extreme eigenvalues are independent is not correct and is removed in this paper. In \cite{Shakir}, when there only exist Gaussian noises, the largest eigenvalue and the smallest eigenvalue of the covariance matrix are assumed to be normal random variables, namely
\begin{subequations}
\begin{align}
&{\lambda _1} \sim{\mathcal{N}}\left( {{u_{{\lambda _1}}},\sigma _{{\lambda _1}}^2} \right)\\
&{\lambda _M} \sim{\mathcal{N}}\left( {{u_{{\lambda _M}}},\sigma _{{\lambda _M}}^2} \right)
\end{align}
\end{subequations}
where $u_{{\lambda _1}}$ and $u_{{\lambda _M}}$ are the means of the largest eigenvalue and of the smallest eigenvalue, respectively. The variances of the largest eigenvalue and of the smallest eigenvalue are denoted by $\sigma _{{\lambda _1}}^2$ and $\sigma _{{\lambda _M}}^2$, respectively. According to \cite{Shakir}, the means and variances of the largest eigenvalue and of the smallest eigenvalue are given as
\begin{subequations}
\begin{align}
&{u_{{\lambda _{\rm K}}}} = \mathbb{E}\left( {{\lambda _{\rm K}}} \right)\\
&\sigma _{{\lambda _{\rm K}}}^2 = \mathbb{E}\left( {\lambda _{\rm K}^2} \right) - u_{{\lambda _{\rm K}}}^2\\
&\mathbb{E}\left( {\lambda _1^p} \right) = C_0^{ - 1}{\beta _{{\lambda _1}}}\left( p \right)\\
&\mathbb{E}\left( {\lambda _M^p} \right) = C_0^{ - 1}{\beta _{{\lambda _M}}}\left( p \right)
\end{align}
\end{subequations}
where ${C_0} = \mathop \Pi \limits_{i = 1}^M \left( {N - i} \right)!\mathop \Pi \limits_{j = 1}^M \left( {M - j} \right)!$; $\mathbb{E}[ \cdot ]$ denotes the expectation operator; $\rm K=\left\{1, M\right\}$; ${\beta _{{\lambda _1}}}\left( p \right)$ and ${\beta _{{\lambda _M}}}\left( p \right)$ are given by eq. $\left(5\right)$ at the top of the next page. In eq. $\left(5\right)$, $\rm sgn\left(\cdot\right)$ denotes the Signum function; $\alpha_{m}$ is the $m$th element of the $\alpha$ and $\alpha$ is the permutation of $\left\{1, 2,\cdots, M-1\right\}$; ${p_{i,j}} = p + N - M + i + j$; $\sum {l_1^{M - 1}}  = \sum\limits_{i = 1}^{M - 1} {{l_i}}$; $l_1^{M - 1}! = \prod\limits_{i = 1}^{M - 1} {{l_i}!}
$; $\sum\nolimits_{{l_{1 \sim M - 1}}}^{{L_{1 \sim M - 1}}} {}  = \sum\nolimits_{{l_1} = 0}^{{L_1}} {\sum\nolimits_{{l_2} = 0}^{{L_2}}  \cdots  } \sum\nolimits_{{l_{M - 1}} = 0}^{{L_{M - 1}}} $; $\mathcal{S}$ is any subset of the set $\left\{l_1, l_2,\cdots, l_{M-1}\right\}$ and $l_m$ is from $0$ to ${L_{{\alpha _m},m}}-1$; ${\left| {\mathcal{S}} \right|}$ represents the cardinality of subset $\mathcal{S}$.
\begin{figure*}[!t]
\normalsize
\begin{subequations}
\begin{align}
&{\beta _{{\lambda _M}}}\left( p \right) = \sum\limits_{i,j}^M {{{\left( { - 1} \right)}^{i + j}}} \sum\limits_\alpha  {{\mathop{\rm sgn}} \left( \alpha  \right)\prod\limits_{m = 1}^{M - 1} {\Gamma \left( {{L_{{\alpha _m},m}}} \right)} } \left( {\sum\nolimits_{{l_{1 \sim M - 1}}}^{{L_{1 \sim M - 1}}} {\frac{{\Gamma \left( {\sum {l_1^{M - 1}}  + {p_{i,j}} - 1} \right)}}{{l_1^{M - 1}!{M^{\sum {l_1^{M - 1}}  + {p_{i,j}} - 1}}}}} } \right)\\
&{\beta _{{\lambda _1}}}\left( p \right) = \sum\limits_{i,j}^M {{{\left( { - 1} \right)}^{i + j}}} \sum\limits_\alpha  {{\mathop{\rm sgn}} \left( \alpha  \right)\prod\limits_{m = 1}^{M - 1} {\Gamma \left( {{L_{{\alpha _m},m}}} \right)} } \left( {\sum\limits_\mathcal{S} {{{\left( { - 1} \right)}^{\left| {\mathcal{S}} \right|}} }\frac{{\Gamma \left( {\sum \mathcal{S } + {p_{i,j}} - 1} \right)}}{{\prod {l_1^{M - 1}!{M^{\sum {l_1^{M - 1}}  + {p_{i,j}} - 1}}} }}} \right)\\
&{L_{{\alpha _m},m}} = \left\{ \begin{array}{l}
\begin{array}{*{20}{c}}
{N - M + m + {\alpha _m} - 1}&{\text{if}\ {\alpha _m} < i\ \text{and} \ m < j}
\end{array}\\
\begin{array}{*{20}{c}}
{N - M + m + {\alpha _m} + 1}&{\text{if}\ {\alpha _m} \ge i\ \text{and}\ m \ge j}
\end{array}\\
\begin{array}{*{20}{c}}
{N - M + m + {\alpha _m}}&{\text{otherwise}}
\end{array}
\end{array} \right.
\end{align}
\end{subequations}
\hrulefill \vspace*{4pt}
\end{figure*}

Since the largest eigenvalue and the smallest eigenvalue are order statistics, the joint PDF, $f_{r,s}\left( {x,y} \right)$, of ${X_r}$ and ${X_s}$, $1 \le r < s \le M$, ${X_r} \le {X_s}$, for $x \le y$, is \cite{H. A. David},
\begin{align} \notag
{f_{r,s}}\left( {x,y} \right) =& \frac{{M!\left[{F_r}\left( x \right)\right]^{r - 1}{f_r}\left( x \right)}{f_s}\left( y \right)}{{\left( {r - 1} \right)!\left( {s - r - 1} \right)!\left( {M - s} \right)!}} \\
&\times{\left[ {{F_s}\left( y \right) - {F_r}\left( x \right)} \right]^{s - r - 1}}{\left[ {1 - {F_s}\left( y \right)} \right]^{M - s}}
\end{align}
where ${F_r}\left( x \right)$ and ${F_s}\left( y \right)$ are the cumulative marginal distribution functions (CMDFs) for ${X_r}$ and ${X_s}$, respectively, and ${f_r}\left( x \right)$ and ${f_s}\left( y \right)$ denote the marginal PDFs for ${X_r}$ and ${X_s}$. Thus, the joint PDF for the largest eigenvalue and the smallest eigenvalue, ${\widetilde f_{{\lambda _1},{\lambda _M}}}\left( {x,y} \right)$, is given by
\begin{align} \notag
{\widetilde f_{{\lambda _1},{\lambda _M}}}\left( {x,y} \right) =& M\left( {M - 1} \right){f_{{\lambda _1}}}\left( x \right){f_{{\lambda _M}}}\left( y \right)\\
&\times{\left[ {{F_{{\lambda _1}}}\left( x \right) - {F_{{\lambda _M}}}\left( y \right)} \right]^{M - 2}}
\end{align}
where $F_{{\lambda _1}}\left( x \right)$ and $F_{{\lambda _M}}\left( y \right)$ are the cumulative distribution functions (CDF) of the two extreme eigenvalues, respectively, and $f_{{\lambda _1}}\left( x \right)$ and $f_{{\lambda _M}}\left( y \right)$ are their corresponding marginal PDFs. $M$ is the number of secondary users. Therefore, the improved PDF for the ratio of the two extreme eigenvalues, ${\widetilde f_Z}\left( z \right)$ is derived as
\begin{align}
{\widetilde f_Z}\left( z \right) = \int_{ - \infty }^\infty  {{{\widetilde f}_{{\lambda _1},{\lambda _M}}}\left( {yz,y} \right)} \left| y \right|dy
\end{align}
where $\left| \cdot \right|$ is the magnitude operator. After substituting eq. $\left(7\right)$ into eq. $\left(8\right)$ and some algebraic manipulations, the improved PDF, ${\widetilde f_Z}\left( z \right)$, is given by
\begin{align}  \notag
{\widetilde f_Z}\left( z \right) = M\left( {M - 1} \right)\sum\limits_{i = 1}^{M - 2} {\left( \begin{array}{l}
M - 2\\
i
\end{array} \right)}{\left( { - 1} \right)^{M - 2 - i}} \\
\times\int_{ - \infty }^\infty  \left[F_{{\lambda _1}}\left( {yz} \right)\right]^i \left[F_{{\lambda _M}}\left( y \right)\right]^{M - 2 - i}{f_{{\lambda _1}}}\left( {yz} \right){f_{{\lambda _M}}}\left( y \right)\left| y \right|dy.
\end{align}
Therefore, the improved PDF for the ratio of the extreme eigenvalues is derived, based on the assumption that the two extreme eigenvalues follow normal distributions \cite{Shakir}, as
\begin{align}  \notag
{\widetilde f_Z}\left( z \right) =& M\left( {M - 1} \right)\sum\limits_{i = 1}^{M - 2} {\left( \begin{array}{l}
M - 2\\
i
\end{array} \right)}{\left( { - 1} \right)^{M - 2 - i}} \\\notag
&\times\int_{ - \infty }^\infty  \Bigg\{\left[{\Phi}\left( {\frac{{yz - {u_1}}}{{{\sigma _1}}}} \right)\right]^i
\left[{\Phi}\left( {\frac{{z - {u_2}}}{{{\sigma _2}}}} \right)\right]^{M - 2 - i}\\
&\times\frac{{\left| y \right|{e^{ - \left[ {\frac{{{{\left( {yz - {u_1}} \right)}^2}}}{{2{\sigma _1}}} + \frac{{{{\left( {y - {u_2}} \right)}^2}}}{{2{\sigma _2}}}} \right]}}}}{{2\pi {\sigma _1}{\sigma _2}}}\Bigg\}dy
\end{align}
where $\Phi\left(x\right)$ is the CDF of the standard normal distribution. Although the solution for the ratio of the two limiting eigenvalues given in eq. $\left(10\right)$ is in integral form, the integral is well behaved, having a stictly positive integrand, and it can be evaluated readily by using standard numerical computation or commonly available mathematical softwares. It is seen that the proposed solution for the PDF of the ratio of the two limiting eigenvalues is simpler than the solution given in [8, eq. $\left(7\right)$]. It is seen from eq. $\left(10\right)$ and eq. $\left(7\right)$ given in \cite{F. Penna2} that the complexity of these two expressions mainly depends on the multiple integration. Moreover, double integrations are required. In \cite{F. Penna2}, there are ${\rm O}\left( {{N^2}} \right)$ additions due to the permutation operation and ${\rm O}\left( {{NM^4}} \right)$ multiplications in the integrals, where ${\rm O}$ is the big ${\rm O}$ notation \cite{G. H. Golub}. In eq. $\left(10\right)$, $M-1$ additions and ${\rm O}\left( {{M}} \right)$ multiplications in the integrals are required. In the simulation, a comparison of the required time for these two expressions is given to further clarify the superiority of our proposed PDF in term of the complexity.

\section{Simulation Evaluations And Discussion}
In this section, simulation results are given to contrast the proposed simple form for the PDF of the ratio of the two limiting eigenvalues and the expressions for the PDF of that ratio given in \cite{F. Penna}, \cite{ Shakir} and \cite{Matthaiou}. We also present some example results that compare the accuracies of the new and previous theoretical approximations for the PDF of the ratio of the eigenvalues to the exact PDF obtained by simulation. The noises are independent identically distributed Gaussian real noises with mean zero and unit variance. All the simulation results are achieved by using $10^6$ Monte Carlo simulations. The number of samples and the number of secondary users are set as $N=50$ and $M=10$ or $N=50$ and $M=20$, respectively.

\begin{figure}[!t]
\centering
\includegraphics[width=3.0 in]{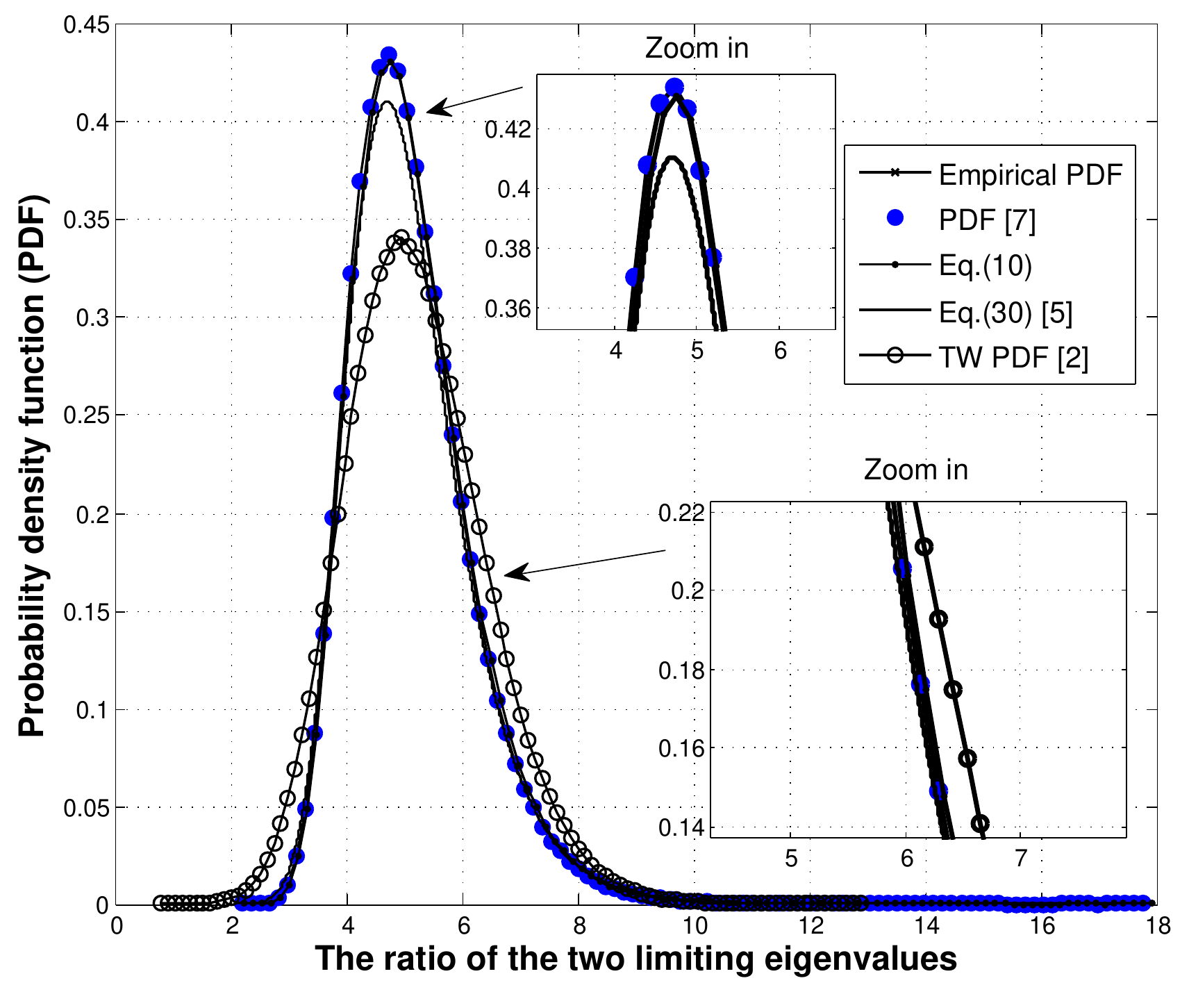}
\caption{Comparison of the improved PDF approximation for the ratio of the two limiting eigenvalues of the covariance matrix with the known approximate PDFs for $N=50$ and $M=10$. } \label{fig.1}
\end{figure}

Fig. 2 shows the commonly employed approximations to the PDF of the ratio of the two limiting eigenvalues of the covariance matrix obtained by using existing methods and our proposed approximation which is obtained without the assumption that the two eigenvalues are independent. In Fig. 2, the empirical PDF curve is the empirical PDF of the ratio of the two liming eigenvalues while the TW PDF is the PDF approximation obtained by using the Tracy-Widom distribution of order $2$ \cite{F. Penna}. The PDF curve labeled eq. (30) [5] is the approximate solution given in [5, eq. $\left(30\right)$]. The PDF curve labeled PDF [7] is the exact PDF given in \cite{Matthaiou}. It is observed that the PDF of the ratio test statistic obtained by using the new solution matches better with the empirical PDF than the PDFs obtained from the other two methods. It is also seen that the PDF given by the exact solution in \cite{Matthaiou} matches well with the empirical PDF. The result consists with the result obtained in \cite{Matthaiou}. The PDF from [5, eq. $\left(30\right)$] and the PDF used to approximate the Tracy-Widom PDF in \cite{F. Penna} are both approximate PDFs, and both are inferior approximations to the new approximation. The Tracy-Widom PDF of order $2$ is not a good approximation to the precise PDF obtained by simulation. The reason is that the Tracy-Widom PDF of order $2$ for the ratio of the two limiting eigenvalues is valid when $\mathop {\lim }\limits_{N \to \infty } \frac{M}{N} = c$, where $c$ is a constant. However, in the simulation, $N$ and $M$ are set as $50$ and $10$, respectively, and these values do not satisfy the limiting condition. It is seen that the new solution provides very high accuracy. The reason is that our new solution is derived without the independence assumption that the samples are independent, and the only source of discrepancy is the Gaussian distribution assumption, which causes only small discrepancies when the number of samples is even moderately large. These results show that the major source of error in the previous approximations is the independence assumption, and not the Gaussian assumption.

\begin{figure}[!t]
\centering
\includegraphics[width=3.0 in]{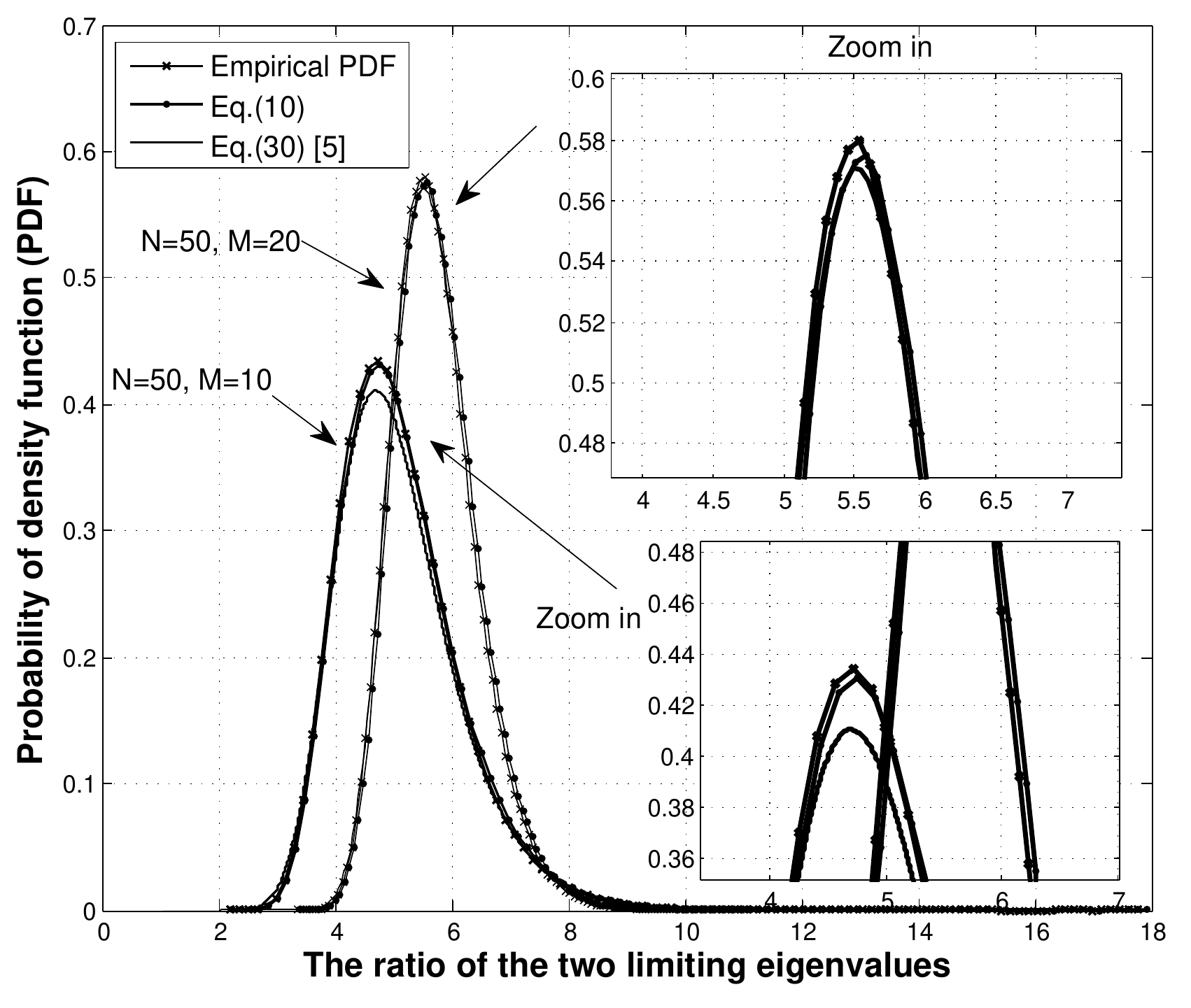}
\caption{Comparison of the improved PDF approximation for the ratio of the two limiting eigenvalues of the covariance matrix with the known approximate PDFs, for $N = 50$ and $M = 10$ or $N = 50$ and $M = 20$.} \label{fig.1}
\end{figure}
Fig. 3 shows the comparison of the improved PDF approximation for the ratio of the two limiting eigenvalues of the covariance matrix with the known approximate PDF given by eq. (30) [5] with different  $M$. It is seen that the accuracy of both our proposed solution and the form given by [5] increases with $M$. The reason is that the accuracy of the mean and variance of the two limiting eigenvalues increases with $M$.

\begin{table}
\begin{center}
\caption{Comparison of  the required computation times (s)}
\begin{tabular}{|c|c|c|c|c|}\hline
\backslashbox{Schemes}{$ \left(N, M\right)$}          &$\left(50, 5\right)$ &$\left(100, 5\right)$ &$\left(100, 10\right)$ &$\left(100, 20\right)$\\\hline
Eq. $\left(7\right)$ in \cite{F. Penna2} &10.248 &14.835&27.482&43.498 \\\hline
Eq. $\left(10\right)$ &6.529&6.572&10.593&22.179 \\\hline
\end{tabular}
\end{center}
\end{table}
In order to compare the complexity of our proposed PDF form with that of the form given by eq. $\left(7\right)$ in \cite{F. Penna2}, the computation times for different parameters $\left(N, M\right)$ are given in Table 1. The results are obtained by using a computer with 64-bit Intel(R) Core(TM) i7-4790 CPU, $8$ GB RAM. It is seen from Table 1 that the required time for calculating the PDF given by eq. $\left(7\right)$ in \cite{F. Penna2} is larger than that for our proposed PDF. This indicates the complexity of our proposed expression is lower than that presented in \cite{F. Penna2}. It further verifies that our proposed PDF is simpler than that proposed in \cite{F. Penna2}.


\section{Conclusion}
A new approximation for the PDF of the ratio of the two limiting eigenvalues of the covariance matrix in eigenvalue-based spectrum sensing was derived based on order statistic analysis. The new approximate solution is the most accurate approximation known, and its derivation does not rely on an invalid independence assumption used to derive a popular previous approximation.  The precise new approximation was used to show that the major source of error in previous approximation is the independence assumption and not Gaussian approximation.
The relative poorness of the Tracy-Widom approximation was clarified and explained.

\end{document}